\documentclass[letterpaper]{article}
\usepackage{aaai}
\usepackage{times}  
\usepackage{helvet} 
\usepackage{courier}  
\usepackage[hyphens]{url}  
\usepackage{graphicx} 
\usepackage{wrapfig}
\usepackage{booktabs}
\usepackage{subcaption}
\usepackage{textcomp}
\usepackage{comment}
\usepackage{multirow}
\usepackage{hyperref}
\usepackage{fixbib}

\usepackage{graphicx}  
\begin{document}

 \setcounter{secnumdepth}{0} 
%

\title{COVID-19 in Spain and India: Comparing Policy Implications by Analyzing Epidemiological and Social Media Data}

\author{\normalsize Parth Asawa,\textsuperscript{1}
Manas Gaur,\textsuperscript{2}
Kaushik Roy, \textsuperscript{2}
Amit Sheth \textsuperscript{2} \\

\textsuperscript{1}{Monta Vista High School, Cupertino, CA, USA  \\ pgasawa@gmail.com} \\
\textsuperscript{2}{Artificial Intelligence Institute, University of South Carolina, Columbia, SC, USA \\ mgaur@email.sc.edu, kaushikr@email.sc.edu, amit@sc.edu}}

\maketitle

\begin{abstract}
\begin{quote}
The COVID-19 pandemic has forced public health experts to develop contingent policies to stem the spread of infection, including measures such as partial/complete lockdowns. The effectiveness of these policies has varied with geography, population distribution, and effectiveness in implementation. Consequently, some nations (e.g., Taiwan, Haiti) have been more successful than others (e.g., United States) in curbing the outbreak. A data-driven investigation into effective public health policies of a country would allow public health experts in other nations to decide future courses of action to control the outbreaks of disease and epidemics. We chose Spain and India to present our analysis on regions that were similar in terms of certain factors: (1) population density, (2) unemployment rate, (3) tourism, and (4) quality of living. We posit that citizen ideology obtainable from twitter conversations can provide insights into conformity to policy and suitably reflect on future case predictions. A milestone when the curves show the number of new cases diverging from each other is used to define a time period to extract policy-related tweets while the concepts from a causality network of policy-dependent sub-events are used to generate concept clouds. The number of new cases is predicted using sentiment scores in a regression model. We see that the new case predictions reflects twitter sentiment, meaningfully tied to a trigger sub-event that enables policy-related findings for Spain and India to be effectively compared.
\end{quote}
\end{abstract}

\section{Introduction}
\begin{figure}[t]
\centering
    \includegraphics[width=60mm,scale=0.5, angle=-90]{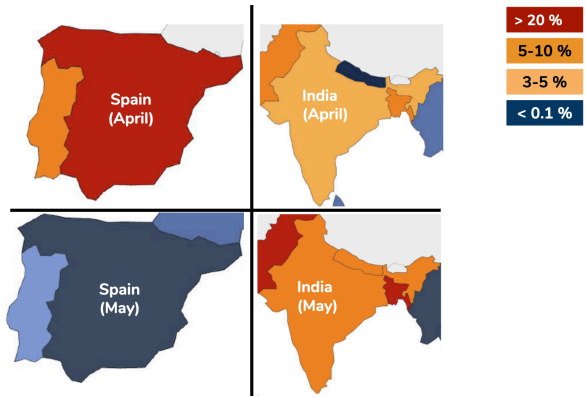}
  \caption{Top Row: April 11th, Spain 27.8\% and India 4.3\%, where x\% refers to the share of COVID-19 tests that came back as positive in a 7-day rolling average. Bottom Row: June 5th, Spain 0.9\% and India 6.7\%, where x\% refers to the share of COVID-19 tests that came back as positive in a 7-day rolling average.}
  \label{fig:1}
  \vspace{-1.75em}
\end{figure}

\begin{figure}[h!]
    \centering
    \includegraphics[width=0.5\textwidth]{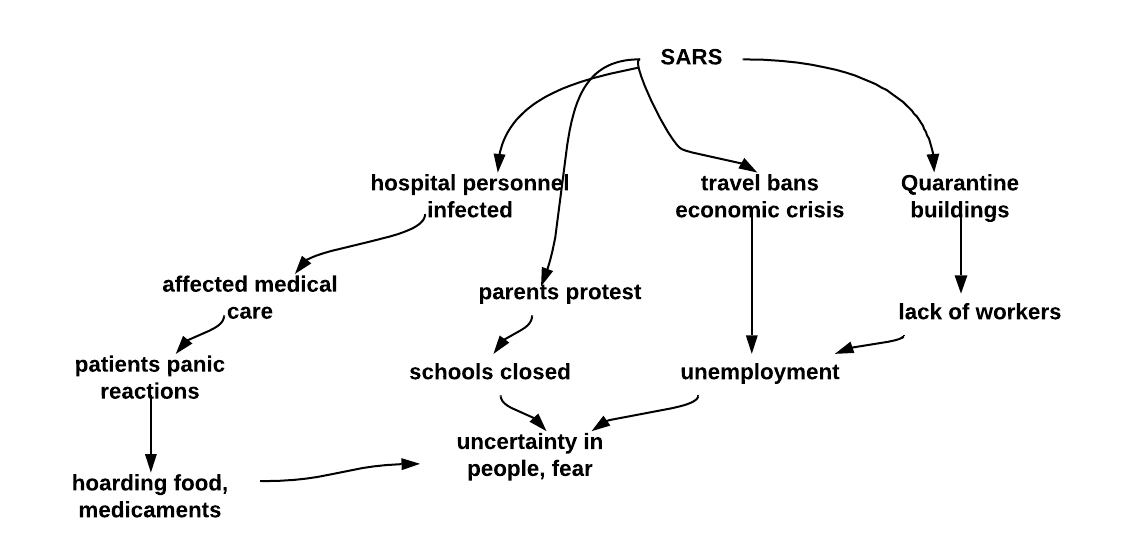}
    \caption{Causality network of sub-events during SARS Pandemic by Helbing et al.. We utilized this graph to represent sub-events within the COVID-19 pandemic during extraction of the word cloud}
    \label{fig:8}
\end{figure}

The COVID-19 pandemic has seen several countries become epicenters for spread. Spain was one such country; however, their policies were effective in curbing the initial outbreak of COVID-19 in March-May of 2020. This is arguably due to people and governments taking precautions to limit the population of people susceptible to the virus — masks, social distancing, lockdowns, business closures, etc from an early stage\footnote{\url{https://www.healthaffairs.org/doi/10.1377/hlthaff.2020.00818}}. Accordingly, the effectiveness of individual countries' policy responses to an epidemic or pandemic can be determined by how well citizens respond to those policies\footnote{\url{http://bit.ly/citizenResponses}}. A person's conformity to a policy may be inferred from their ideologies mined through social media, such as Twitter \cite{van2020impact}. As shown in figure \ref{fig:1}, over three months, Spain recorded a decline of 97\% in the number of new cases, whereas India has shown a 36\% influx in new patients. Is it possible to explore policy transfer from Spain to India to curb the alarming COVID-19 cases? Could the number of infections be modeled using the Twitter concepts about causal trigger sub-events in a causality network \cite{helbing2006disasters}? The reason we are conducting this study is there is limited prior research relating policy and changes in case counts, through social media analysis, for COVID-19.
We use Twitter as the active platform for live information on the spread of COVID-19. Government policies, especially in developing nations, based on the epidemiological data, ignore the population-specific behaviors of culture, ideology, and politics that hinder these policies' implementation. For example, a large number of people in the US are opposed to wearing masks. To this end, we juxtapose Spain and India's epidemiological data to identify a date when the curves show the number of new cases diverging from each other, and India started showing worsening conditions.Although it could be argued that the differences we see in cases were due to travel from hotspots, it's important to note that India closed its borders by suspending all international flights starting March 22nd, in addition to taking steps to suspend inter-state travel by suspending domestic flights and domestic trains throughout the time frame of our analysis\footnote{\url{https://www.nytimes.com/article/coronavirus-travel-restrictions.html}}. We recognized some critical policy-related concepts which are causally related in the COVID-19 context. For instance, ``settlement areas'', ``confinement to barracks'', ``mistrust of people'', ``loss of government authority'' causally follow announcement of ``public policy''. Hence, we used the causality network of policy-related concepts identified by experts during severe acute respiratory syndrome (SARS) to perform a  knowledge-guided search on Twitter \cite{helbing2006disasters} (see Figure \ref{fig:8}). We show Kerala and Mumbai's policy-related concept clouds. Then we investigate the applicability of interventional policies in Madrid and Barcelona to Kerala and Mumbai. Likewise, we observed a policy-level association between the Canary Islands and Andhra Pradesh as both regions have strong healthcare infrastructure. \\ \indent The main contributions of this work are thus investigating Twitter conversations corresponding to explanatory causal trigger events, to form an ideological map of the population that provides insights into response to government policy (see Methods). In turn, this is validated through the prediction of new cases using the sentiment scores of the twitter conversation (see Regression Analysis and Explanatory events). Finally, a comparison of policy and responses across similar regions in Spain and India is discussed (see Discussion and Findings).

\section{Related Work}
\cite{cowling2020impact} statistically analyzed the impact of policy on reducing the transmissibility rate of COVID-19. The study was conducted on the epidemiological data of Hong Kong, and inferences were made using confidence intervals. Our research aims to investigate the applicability of policies created by developed nations onto developing nations. Such an exploration is not possible in Cowling et al.'s study. Further, Cowling et al. provide statistical explanations on government policies' potency in Hong Kong rather than conceptual explanations, which is required to decide the ``what next.'' While probing government policies' relevance from one nation to another, population-specific behaviors negatively affect cross-nation policy transfer. For instance, a likely source of infection in India was the Tablighi Jamaat movement, a religious gathering \footnote{\url{https://www.aljazeera.com/news/2020/04/tablighi-jamaat-event-india-worst-coronavirus-vector-200407052957511.html}}, which became a coronavirus vector and was not taken into account in government policy or enforcement \cite{sivaraman2020exo}. Likewise, the return of migrant laborers to their home states in India and long weekend celebrations and parties in the United States led to an increase in COVID-19 cases. As a result, policies such as reopening, contact tracing, and ensuring public compliance, which was effective in Europe, are not directly applicable to India and the United States \cite{hellewell2020feasibility}. It is essential to relate patterns in epidemiological data with evolving policy-related concepts and sentiment on social media to better study the likelihood of policy effectiveness \cite{Kalteh2020}. Other regression models that predict new cases do not consider social media information, which we posit is a significant predictor \cite{shayaknew} \cite{prem2020effect}.

\section{Materials and Methods}
\subsection{Materials}

In this research problem, we use multiple publicly available datasets and government resources, specific to Spain and India (e.g., news reports, insights on epidemiological data).

The first country dataset is a COVID-19 dataset for Spain data. The dataset is available here: \href{https://github.com/victorvicpal/COVID19_es}{\textbf{Link}}. It contains attributes including but not limited to: 
Total \# of Cases, Total \# of Hospitalizations, Total \# of Patients in the ICU, Total \# of Recovered Patients, and Total \# of New Cases. The dataset was derived entirely from Spain's Ministry of Health website and transformed into CSV files.  All of the data is available by province (the equivalent to states in the United States). The second dataset we use is a COVID-19 dataset for India, available here\footnote{\url{https://api.covid19india.org/}}. This dataset contains attributes including but not limited to:  \# of Confirmed Cases per Day, \# of Recovered per Day, \# of Deaths per Day, \# of People in the ICU, \# of People on Ventilators. The dataset was sourced from several sources, a list of which can be found here\footnote{\url{https://telegra.ph/Covid-19-Sources-03-19}}. All of the data is available on a state-by-state level within India. \\ \indent After having the two datasets for identifying divergence points and initial identification of a problem, the final dataset we use is a dataset of Twitter-IDs, for our twitter social media analysis available here\footnote{\url{https://github.com/echen102/COVID-19-TweetIDs}}.  As stated in the dataset, "The repository contains an ongoing collection of tweets IDs associated with the novel coronavirus COVID-19 (SARS-CoV-2), which commenced on January 28, 2020. We used Twitter's search API to gather historical Tweets from the preceding seven days, leading to the first Tweets in our dataset dating back to January 21, 2020." This dataset gives us access to the Tweet ID's pre-filtered concerning the coronavirus with keywords accessible here\footnote{\url{https://github.com/echen102/COVID-19-TweetIDs/blob/master/keywords.txt}}. From this dataset, we hydrated $5,075,830$ tweets from April 15 to May 15, of which $534$ were geotagged from the state of Kerala, and $7094$, the state of Mumbai. 

\subsection{Methods}

\begin{figure*}[t]
\centering
    \includegraphics[width=0.8\textwidth]{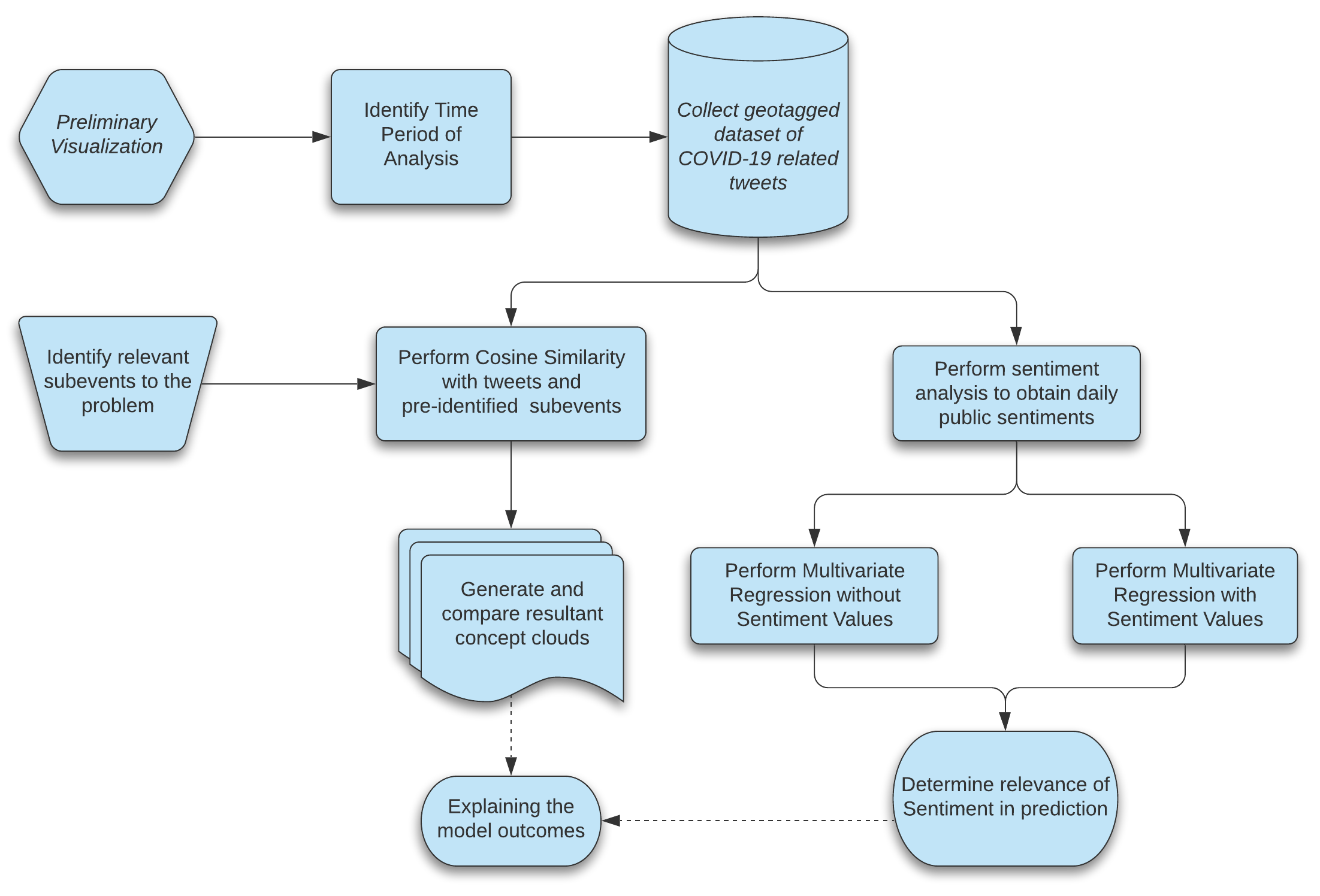}
  \caption{Workflow detailing the approach described in this study to analyze citizen response to policies and generate explainable inferences on the epidemiological data, in addition to predicting future changes in the spread of an epidemic.}
  \label{fig:workflow}
  \vspace{-1em}
\end{figure*}

We want to analyze the differences between the spread of the virus in Spain and India; however, the countries are too diverse to compare in their entirety. Thus, we instead propose comparing the two countries on more granular scales, specifically by identifying pairs of states/regions (India/Spain) that are similar on the following grounds: (1) population density, (2) unemployment rate, (3) tourism, and (4) quality of living, and examining the results. For this study, we restrict to the following two pairs of states/regions: (1) Kerala and Madrid, and (2) Maharastra (Mumbai city) and Catalu$\tilde{n}$a (Barcelona region). 

\begin{figure}[ht]
\centering
    \includegraphics[width=60mm,scale=0.5, angle=-90]{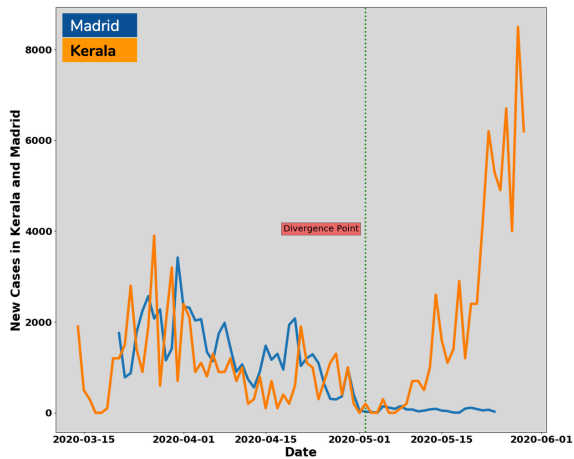}
  \caption{ Daily New Cases of COVID-19 in Kerala (scaled up by 100 for visibility) and Madrid plotted against time from March 15th to June 1st, with an identified Divergence Point of where the two curves no longer follow the same trend.}
  \label{fig:2}
  \vspace{-1em}
\end{figure}

\begin{figure}[ht]
    \includegraphics[width=60mm,scale=0.5, angle=-90]{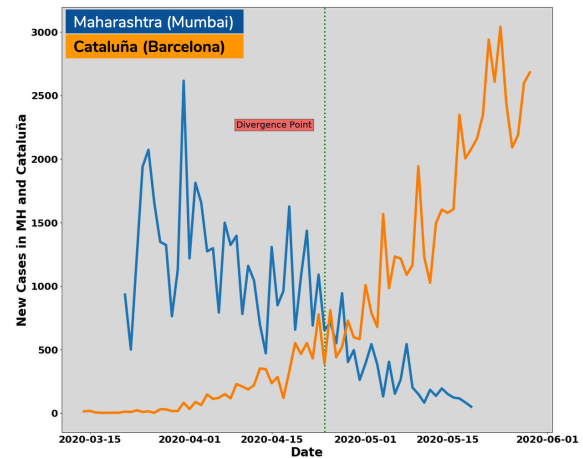}
  \caption{Daily New Cases of COVID-19 in Maharashtra (Mumbai City) and Catalu$\tilde{n}$a (Barcelona region) plotted against time from March 15th to June 1st, with an identified Divergence Point of where the two curves intersected.}
  \label{fig:3}
  \vspace{-1.5em}

\end{figure}

On the data from these states/regions, we did visualizations of counts of new cases during April and May. This period was essential to assess the effectiveness of government policies in controlling the COVID-19 pandemic. By creating pairs of states/regions from India and Spain, we identified divergence points where India started showing worsening public health. Figure \ref{fig:2} shows May 1st, 2020, as the divergence point for Kerala and Madrid. Likewise, April 22nd, 2020, is the divergence point for Mumbai and Barcelona (Figure \ref{fig:3}). 

Once the relevant timeframe is defined, we extract tweets geotagged to the local Indian regions, such as Kerala and Mumbai. It allows us to explore the people's responses towards government policies, which helps assess the rise in COVID-19 cases.  Semantically understanding people's reactions from their twitter conversations is a challenging task for statistical natural language processing. Hence, we utilize a hypothesized causal graph of policy-dependent sub-events in Helbing et al., which describes a series of activities occurring during a pandemic. Some of the concepts described by Helbing et al. are mistrust, church hospitals, mask distribution, mental health. We identify a set of relevant concepts that describe Kerala and Mumbai's tweets using a pre-trained multilingual ConceptNet model from a Sem-Eval task \cite{speer2017conceptnet}. We use the Spacy parser to generate phrase embeddings of concepts and nouns extracted from tweets\footnote{\url{https://spacy.io/api/dependencyparser}}. Next, we perform a cosine similarity between the tweet vector and concept vector, with an empirically determined threshold of 0.45. The frequency of concept phrases was recorded and presented as people's responses in the given region during the given time frame.

\section{Exploratory Data Analysis}
\begin{figure}[ht]
\centering
    \includegraphics[width=50mm,scale=0.5, angle=-90, trim=3.0cm 1.0cm 3.0cm 1.0cm]{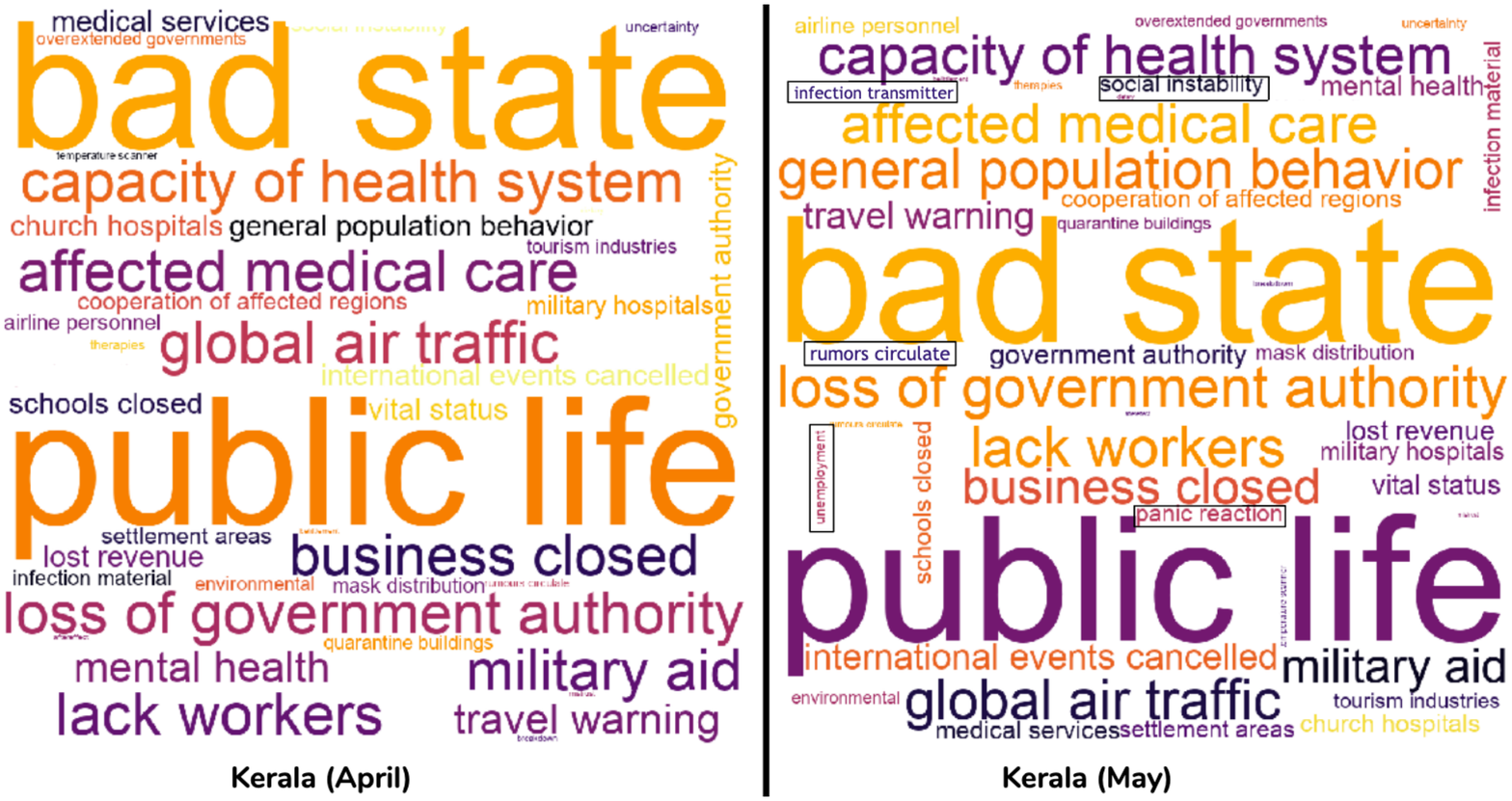}
  \caption{After the first wave of COVID-19 spread in the month of March, the government of India instituted various policies, such as school closings, business closings, travel bans, over-extensions,  which impacted public life, especially for daily wage families. Hence, we see rise in the frequency of tweets concerning mental health, medical care, and unemployment. As a consequence of the policies, we observe emerging events such as rumors, churches becoming hospitals due to overloaded healthcare facilities, social instability, and mistrust (in rectangle black box). Through citizen sensing around the point of inflection (Figure \ref{fig:2}) , we noticed a constant frequency of concepts such as poor public life and bad condition of the state, which reflected on the imperfection in policy implementation.}
  \label{fig:kerala}
  \vspace{-1.5em}
\end{figure}

We begin by performing a preliminary visualization of the dataset. In Figure \ref{fig:2}, we observe the new case counts in Kerala scaled up by a factor of 100 (for trend visibility) compared to Madrid's region. It seems that the data points remained reasonably close from the period of March 15th to May 1st, after which there is a second wave of COVID-19 spread in Kerala. In contrast, Madrid remained relatively close to 0 for the rest of the period. This divergence from its previous relative similarity to Madrid is a key feature we intend to explore using real-time conversations on twitter.  Through semantic analysis of Kerala's tweets around the point of inflection, we recorded mentions of \textit{gatherings} such as marriages and \textit{poor capacity of the health system}, which are potential causes of the rise in new cases (see Figure \ref{fig:kerala}).

Furthermore, people mentioned information on \textit{ways of transmission} with no known source of origin, prompting the government to reinstate lockdown procedures. \textit{Overextension of lockdown by the government} developed a \textit{panic reaction} among the individuals in Kerala. The state also saw a \textit{lack of cooperation among authorities in affected regions}, which contributed to a surge in cases. \textit{Rumors circulated} through misleading campaigns that developed uncertainty and fear upsetting people's livelihood in Kerala, making them restless in critical \textit{containment zones}. From April to May, people's responses to government policies showed expressions of \textit{social instability}, \textit{unemployment}, \textit{uncontrolled infection transmission}, and \textit{circulation of rumors}.

\begin{figure}[ht]
\centering
    \includegraphics[width=50mm,scale=0.5, angle=-90, trim=3.0cm 1.0cm 3.0cm 1.0cm]{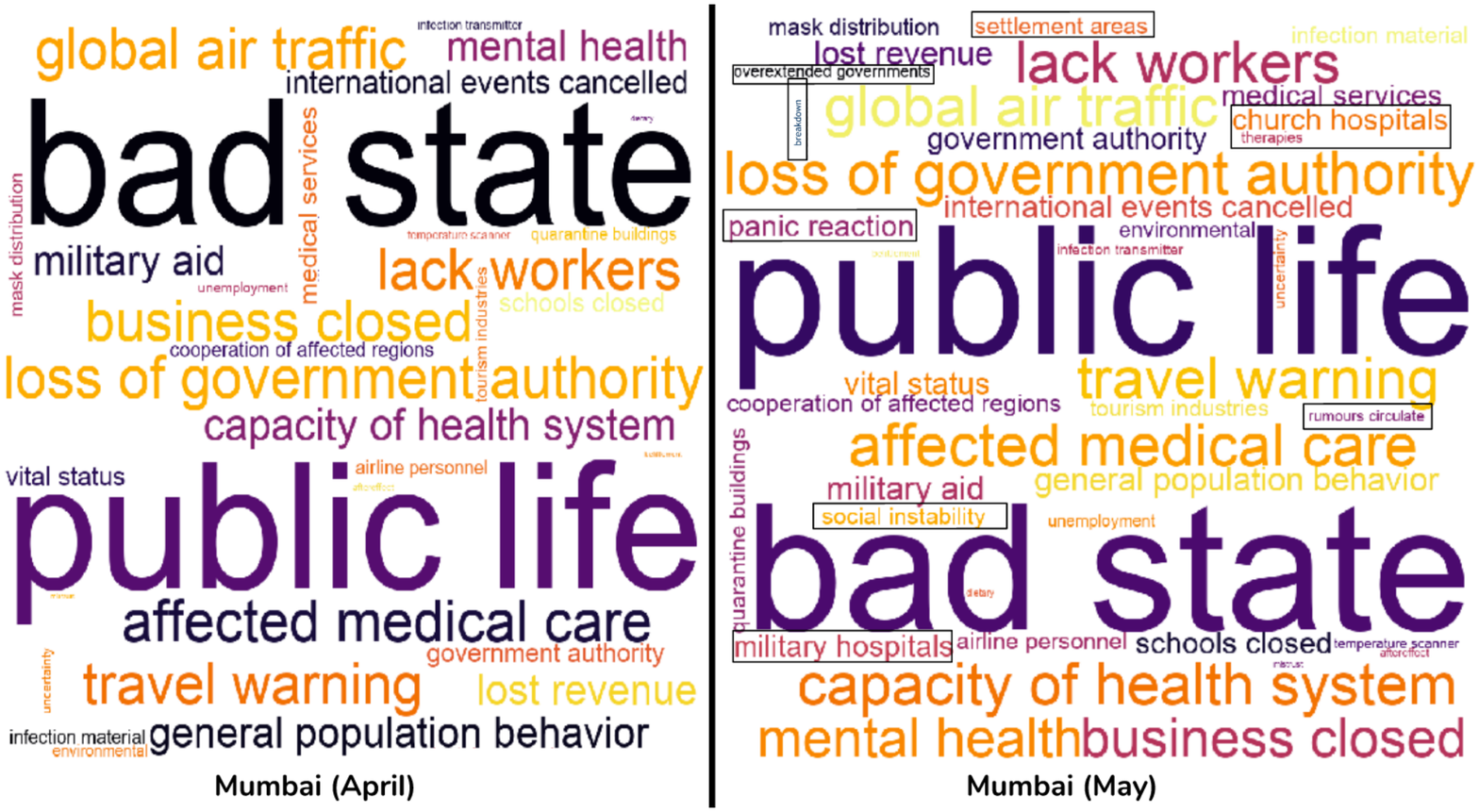}
  \caption{As we can see, within both states, the topical content being discussed is relatively the same. Throughout the frame of the time series, including April, we saw that the trend in coronavirus cases had a steadily increasing number of new cases per day, or a positive second derivative. This indicates that the similarity in thinking over time compounded, possibly resulting in the eventual seemingly exponential growth in the spread of COVID-19 that we witness.}
  \label{fig:mumbai}
  \vspace{-1em}
\end{figure}

In Figure \ref{fig:3}, we observe the plots of daily new cases in Maharashtra, whose case counts were almost all from Mumbai and Catalu$\tilde{n}$a (Spain, Barcelona). First, it seems that the data points remained fairly close from March 15th to April 22nd, at which point the new cases in Catalu$\tilde{n}$a remained fairly close to 0 for the rest of the period. Though the population density and social composition of Mumbai are different from Kerala, we recorded the use of similar concept phrases reflecting similar consequences of government policies. For instance, social instability, reaching out to catholic hospitals\footnote{\url{https://www.licas.news/2020/06/18/as-indias-healthcare-system-struggles-with-covid-19-catholic-hospitals-join-the-front-line/}} (or church hospitals), seeking military aid during lockdown\footnote{\url{https://www.thehindu.com/news/cities/mumbai/lockdown-state-seeks-armys-help/article31188053.ece}}, mental health, panic reaction, and people seeking therapy.  Compared to Kerala, Mumbai showed a significant rise in unemployment, which is relatively similar to the trend in unemployment in Barcelona, and  Madrid\footnote{\url{https://www.theolivepress.es/spain-news/2020/05/12/madrid-and-barcelona-both-rank-in-the-bottom-10-of-best-cities-for-jobs-following-coronavirus-crisis/}}. The situation of unemployment remained constant from April to May in Kerala and Mumbai.
Further, the concept of "general population behavior" describes the migrant population, which constituted 93\% workforce in India, contributed to the rise in the COVID-19 cases as people travelled back to their homes for security. These external factors, which aren't recorded in epidemiological data but explain epidemiology patterns, should be incorporated in models like SIR to better estimate the future patterns in the spread of disease \cite{sivaraman2020exo}.  As we can see, within both states, the topical content being discussed is relatively the same. In the time series curve, including April, we saw that the coronavirus cases had a steadily increasing number of new cases per day with a slight curvature. This indicates that the similarity in thinking over time compounded, possibly resulting in the eventual seemingly exponential growth in the spread of COVID-19. We will next validate if these thinking patterns captured in Twitter sentiments are a good predictor of new cases.

\subsection{Regression Analysis and Explanatory events}
We use Multivariate Linear Regression (MVR) with tweet sentiment to predict future cases in Kerala and Mumbai's regions from mid-April to mid-May, over a month across different periods. To determine each tweet's sentiment, we use the flairNLP Python library \footnote{\url{https://github.com/flairNLP/flair}}. We combine sentiments of concepts (Figure \ref{fig:kerala} and \ref{fig:mumbai}) identified from each tweet into daily sentiment values -- from the period of April 16th to May 14th/15th. We then perform MVR using the features is described in materials sections and another with tweet sentiment. The first MVR model uses the past 30 days of new cases and recovered cases to predict the next 30, and the second MVR model also uses tweet sentiment to predict the next 30 days. We use a cumulative function on both new cases and recovered cases to better reflect the upward trend. \\ \indent We find that the Regression error does indeed decrease when using the tweet sentiments. We specifically look at the differences in the RMSE values and the adjusted $R^2$ for quantitative performance gains. Further, we use periods of $3, 7,$ and $14$ days from May 15th for the two MVR models, as these have been shown in \cite{pavlicek2020oscillatory} to be the periods of days with which COVID-19 deaths show regularities (see Table \ref{tab:kerala_predictions} and \ref{tab:mumbai_predictions}). Previous literature suggests that the RMSE uncertainty for this number of data points would be approximately $12.9\%$ \cite{FABER199979}.

\begin{table}[ht]
\footnotesize
    \centering
    \begin{tabular}{p{2.cm}|p{1.cm}p{1.cm}|p{1.cm}p{1.cm}}
    \toprule[1.5pt]
        \multirow{2}{*}{\textbf{Time period}}& \multicolumn{2}{c}{\textbf{With Sentiment}} & \multicolumn{2}{c}{\textbf{Without Sentiment}} \\ \cmidrule{2-5}
       \textbf{ for Prediction} & RMSE & adj$R^2$ & RMSE & adj$R^2$ \\ \midrule
        14 Days & 9.54 & 0.84 & 11.73 & 0.76 \\
        7 Days & 7.85 & 0.68 & 7.85 & 0.68 \\
        3 Days & 6.46 & 0.63 & 6.51 & 0.63 \\
        \bottomrule[1.5pt]
\end{tabular}
\caption{RMSE and adj$R^2$ Regression Results with and without Sentiment for the State of Kerala, model trained on values from April 16th to May 14th. All the scores are significant with one-tailed t-test at p-value 0.1}
\label{tab:kerala_predictions}
\end{table}

\begin{table}[ht]
\footnotesize
    \centering
    \begin{tabular}{p{2.cm}|p{1.cm}p{1.cm}|p{1.cm}p{1.cm}}
    \toprule[1.5pt]
        \multirow{2}{*}{\textbf{Time period}}& \multicolumn{2}{c}{\textbf{With Sentiment}} & \multicolumn{2}{c}{\textbf{Without Sentiment}} \\ \cmidrule{2-5}
       \textbf{ for Prediction} & RMSE & adj$R^2$ & RMSE & adj$R^2$ \\ \midrule
        14 Days & 286.16 & 0.95 & 310.60 & 0.88 \\
        7 Days & 235.38 & 0.96 & 245.27 & 0.93 \\
        3 Days & 232.09 & 0.97 & 238.57 & 0.92 \\
        \bottomrule[1.5pt]
\end{tabular}
\caption{RMSE and adj$R^2$ Regression Results with and without Sentiment for the State of Mumbai, model trained on values from April 16th to May 14th. All the scores are significant with one-tailed t-test at p-value 0.1}
\label{tab:mumbai_predictions}
\vspace{-1em}
\end{table}
A model's explainability is vital in such a high stakes application for humans to trust and understand its predictions. While the weights of a linear model lend themselves nicely to interpretation, they alone do not provide any insight into the type of events that may have triggered such conversation on Twitter. For tweets with concepts of high sentiment score weight in the model, we use the causal graph \cite{helbing2006disasters} built for the SARS epidemic to provide explanatory sub-event triggers for those concepts. An example is shown in Figure \ref{fig:8}, where the causal structure of sub-events that guided the extraction of twitter conversation is marked. The government can use this graphical explanation to shape its policy going forward.
\\ \indent Note that the dataset of Mumbai tweets was 14 times more extensive than Kerala, resulting in high RMSE. We see a more noticeable difference in adj$R^2$ and RMSE values for Mumbai further in time from May 15th, than we do for Kerala except for the 14 days. Thus, we believe that this research can be explored further with potentially more statistically significant findings through access to larger datasets and more extensive experimentation. However, the increase of the accuracy of using sentiment does seem to happen for both states further away from May 15th, i.e., the model extrapolates better.

\section{Discussion and Findings}
In this paper, we presented a methodology to determine crowd responses to governmental policies that can impact health and new case predictions in real-time, and evaluate those responses to provide direction for new public health policy.

In broad terms, the method presented is the first visualization of the data to identify the features of interest, elicit time-frames of events upon which to focus analysis, and explain the pattern in epidemiological data with social network sentiment analysis. For our comparison of the effectiveness of policies in Spain and India, we were able to identify a critical time-frame across multiple state/province pairs that proved to be a divergence point in the spread of the virus where Spain appeared to be succeeding in containing the virus. In contrast, India seemed to be experiencing exponential growth. Looking at the timelines of government lockdowns: After the 10th case, India took action on Day 21 and Spain on Day 16. After the 1st death, India took action on Day 13 and Spain on Day 29. Finally, after the 100th case, India took action on Day 13 and Spain on Day 10.\\ \indent We see that arguably, the nations took action on a similar timescale concerning the beginning of the spread. We posit, therefore, that the differences in responses to policies can be found in crowd ideology via Twitter. Looking at a few of the previously identified key phrases, we can see some examples of selected tweets that display concepts  previously identified in the concept clouds, along with a timely response from authorities in Spain:

\begin{enumerate}
    \item Tourism tweet (Kerala): ``\textit{One of the largest sectors of \#Indianeconomy, \#Tourism, lies in tatters due to the \#CoronaPandemic and the \#lockdown}''----- Spain chose to handle tourism by closing its border to outsiders, as of April, only allowing diplomats, traveling for emergencies, or residents of the European Union, and assorted smaller states\footnote{\url{https://www.euronews.com/2020/05/23/spain-will-open-borders-to-foreign-tourists-in-july-in-phasing-out-of-coronavirus-restrict}}.
    \item Medical Care tweet (Mumbai): ``\textit{When the richest country has zero public health care in place and they need to hire in the middle of a pandemic}'' ----- Spain used a royal decree to declare a 15-day national emergency back on March 15th \cite{legido2020resilience}. It dedicated significant investments to its healthcare system, quoted ``It had allocated \texteuro 2.8 billion to all regions for health services and created a new fund with \texteuro 1 billion for priority health interventions.''
    \item Social Instability tweets (Kerala and Mumbai): (a) ``\textit{If you get into a cyclical lockdown it will be devastating for economic activity because that would destroy trust.}''(b) ``\textit{People will lose trust if the lockdown continues indefinitely. Need to work out a way.\#RahulShowsTheWay}'' ----- Spain's Civil Guard dedicated time to compiling a report and evaluating possible scenarios of growing social unrest in conjunction with law enforcement agencies, coming up with different responses to rising crime rates or civil unrest. The report specifically noted that the Spanish population has accepted the lockdown, “which started out as one of the strictest in Europe” \footnote{\url{https://english.elpais.com/society/2020-05-15/spains-civil-guard-warns-about-risk-of-social-unrest-due-to-covid-19-crisis.html}}. 
    \item Cancelled Events tweets (Mumbai): ``\textit{\#MAMI Mumbai Film Festival 2020 cancelled. Second major event in Mumbai to be cancelled this year after Lalbaugcha Raja Ganeshotsav. Cannot imagine the loss of revenues.}'' ----- A number of events, such as Easter Sunday, were cancelled in Spain\footnote{\url{https://gulfnews.com/world/europe/easter-sunday-events-in-spain-cancelled-communities-make-masks-amid-virus-outbreak-1.1586627331285}}. Further, a selective set of interntional events were allowed with limited capacity and stringent laws (e.g. Live Music) \footnote{\url{https://www.nme.com/news/music/spain-to-phase-in-live-music-events-in-may-as-part-of-lockdown-exit-plan-2656841}}.

\end{enumerate}

This is where real-time NLP analysis plays an instrumental role. Identifying topical categories and sentiments associated with them through social network analyses like Twitter provides an avenue to quantitatively and qualitatively evaluate and rank responses to different policies. For quantitative assessment, we considered intuitive model performance metrics, such as RMSE and adj$R^2$. Qualitative inspection was performed by mapping the people's response to sub-events in SARS's causality network. We project the identified causally triggered sub-events onto a concept cloud and analyze over two critical months post-initiation policies. Even though a linear model is already interpretable in terms of weights, this type of explainability is of paramount importance to understand and trust the model predictions in such a high stakes application. This can give governments insight into whether they must make policies stricter, add more policies, or enforce policies differently than they are at the moment. Real-time analysis of the social network and virus data can significantly change the course of health events and are a promising yet relatively unexplored tool for governments and policymakers to use.

\section{Future Work}
\begin{figure}[!ht]
\centering
    \includegraphics[width=60mm,scale=0.5, angle=-90]{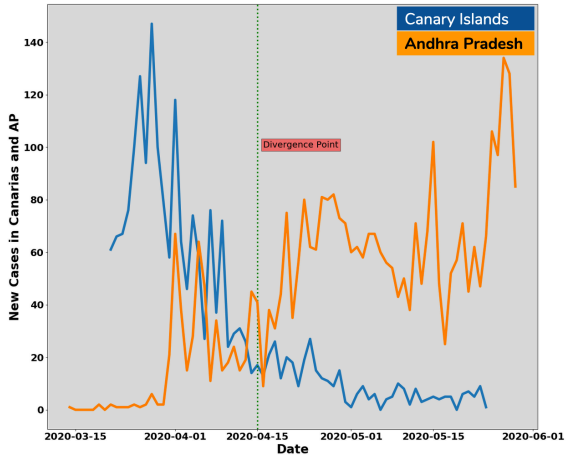}
  \caption{Daily New Cases of COVID-19 in Andhra Pradesh (not scaled) and the Canary Islands plotted against time from March 15th to June 1st, with an identified Divergence Point of where the two curves intersected.}
  \label{fig:4}
  \vspace{-1em}
\end{figure}

We have presented in this work a case study with two (State, Region) pairs, specifically (Mumbai, Barcelona) and (Kerala, Madrid). We posit that this work can be extended to other (State, County) pairs. Considering one pair such as Andhra Pradesh and the Canary Islands (see Figure \ref{fig:4}) — both of which are known to have strong healthcare systems relative to the rest of their countries — we can plot the time series visualization and analyze the divergence point.

It's important to note that there other uncontrolled variables that make it hard to draw affirmative causal conclusions, and this is an important aspect we hope to consider in future work. The results from this preliminary work could be used to explain epidemiological models, specifically, the Exo-SIR (Exogenous - Susceptible, Infected, Recovered) model. Exo-SIR is built to model the disease's spread while taking into account exogenous factors (e.g., gathering, compliance to public policy). Since our study identified concepts such as social instability, mistrust, and poor medicare as responses of the population against the instated policies, it could be considered potential exogenous factors influencing SIR models. Our future research may entail including government policies themselves as the Exogenous impact on a SIR population, and more accurately identifying and explaining the spread of a disease in a community by considering citizen response to policies.

All the code and datasets for this study are available for the reproducibility of our results \href{https://github.com/pgasawa/AIISC_SpainIndia_COVID-19_Modeling}{here}.
\section{Acknowledgements}
We would like to acknowledge Dr. Victor Vicente Palacios for his support in the Spain data collection and its interpretation. Also we would like to acknowledge Mr. Nirmal Sivaraman and Dr. Sakthi Balan of LNMIIT-Jaipur for their brainstorming and input into the direction of this research. We acknowledge partial support from the National Science Foundation (NSF) award 1761880: "Spokes: MEDIUM: MIDWEST: Collaborative: Community-Driven Data Engineering for Substance Abuse Prevention in the Rural Midwest". Any opinions, conclusions or recommendations expressed in this material are those of the authors and do not necessarily reflect the views of the NSF.

\bibliography{references}
\bibliographystyle{aaai}

\end{document}